\documentclass[twocolumn]{revtex4}

\usepackage{epsfig}

\usepackage{amsmath}
\usepackage{amsfonts}
\usepackage{amssymb}

\begin{document}

\title{Cluster approximations for probabilistic systems: \\
a new perspective of epidemiological modelling}
\author{Thomas Petermann$^1$}
\email{Thomas.Petermann@epfl.ch}
\author{Paolo De Los Rios$^{1}$}
\affiliation{$^1$Institut de Physique Th\'eorique, 
Universit\'e de Lausanne, CH-1015, Lausanne, Switzerland.}
\altaffiliation{Present address: Laboratoire de Biophysique Statistique, ITP - FSB, Ecole Polytechnique F\'ed\'erale de Lausanne, CH-1015 Lausanne, Switzerland.}
\date{\today}

\begin{abstract}
Especially in lattice structured populations, homogeneous mixing represents an inadequate assumption. Various improvements upon the ordinary pair approximation based on a number of assumptions concerning the higher-order correlations have been proposed. To find approaches that allow for a derivation of their dynamics remains a great challenge. By representing the population with its connectivity patterns as a homogeneous network, we propose a systematic methodology for the description of the epidemic dynamics that takes into account spatial correlations up to a desired range. The equations which the dynamical correlations are subject to, are derived in a straightforward way, and they are solved very efficiently due to their binary character. The method embeds very naturally spatial patterns such as the presence of loops characterizing the square lattice or the treelike structure ubiquitous in random networks, providing an improved description of the steady state as well as the invasion dynamics.
\end{abstract}


\maketitle

\section{Introduction}

The spreading dynamics of an infectious disease is crucially determined by the population's underlying connectivity patterns. Followed by the renewed interest in graph theory witnessed by statistical physics in the recent years \cite{albert,dorogovtsev}, substantial progress has been achieved in the field of epidemiology. As an example, the scale-free degree distribution of the Internet accounts for the absence of a finite epidemic threshold in the spreading of a computer virus \cite{pastor,white}. The long persistence of HIV is due to the very same topological property of the web of human sexual contacts \cite{liljeros}, and only targeted immunization leads to a finite epidemic threshold \cite{barabasi}.

Most of these studies are based on the mean-field approximation which represents a reasonable assumption for networks that exhibit large connectivity fluctuations (that is $\langle k^2 \rangle \rightarrow \infty$ where $k$ represents the number of emanating links from a specific node, and the average is taken over the entire network). Quite often however, e.g.\ in lattice structured populations, spatial correlations become important and heterogeneous mixing has to be taken into account. Matsuda et al.\ \cite{matsuda} first used the ordinary pair approximation for the treatment of a biological problem, and the resulting improvements are considerable. In addition to the analytical tractability of this approximation, simple estimates for the epidemic threshold can be obtained in terms of a ``dyad heuristic'': a condition for the location of the critical point is elaborated by looking at two neighbouring infected sites, comparing its recovery with the infections it gives rise to \cite{durrett96}.

Various extensions upon the standard pair approximation have been proposed. In order to analyze the propagation of a wavelike invasion in a lattice structured population, a remarkable improvement is brought about if the region occupied by the infected individuals is described by the ordinary pair approximation, and the leading edge of the wave front is modelled by the quasi-steady-state pair approximation \cite{ellner}. Bauch and Rand developed a pair model for the situation where the population's underlying connectivity patterns are of a dynamical nature \cite{bauch}.

The grid-like structure or the presence of triangles are topological properties which both the mean-field and the standard pair approximations do not account for. An approach pursued by different authors is to use parameters that characterize the topology (such as the density of triangles) and to make a number of assumptions about the corresponding higher-order correlations, which leads to improved pair models. Van Baalen illustrates this method for the triangular and square lattices, if the higher-order correlations are set to 1 \cite{vanbaalen}. The invasion dynamics is reproduced very accurately. The same strategy can be explored for less homogeneous networks \cite{keeling97}, and the consequences regarding epidemiological invasions have been discussed in detail \cite{keeling99}. The improved pair approximation \cite{sato94,sato00} takes into account the clustering property of lattice models more precisely. Its key ingredient is to make less restrictive assumptions about the higher-order correlations, for example they can be set to a value not equal to 1.

Morris derived the dynamics of higher-order correlations in the usual vein, that is an equation which determines the time evolution of an average quantitiy is used as point of departure \cite{morris}.

In this paper, we introduce a novel method for the description of the epidemic dynamics which takes into account spatial correlations up to a desired range. The methodology is illustrated for the case where the population's underlying connectivity patterns is given by a homogeneous network. Concerning the local contact process, we use a susceptible-infected-susceptible model involving transition rates between the two possible states (e.g.\ \cite{durrett94}). The formalism is elaborated in discrete time, and the continuous-time dynamics arises as a limiting case. This limit has been performed in order to allow for a comparison with the above sketched approaches.

The paper is organized as follows. In Section II, the adopted model involving the local dynamics as well as the selected geometries is described in detail. Section III reviews the mean-field and pair approximations, introduces our formalism and explains how these approximations are recovered. In Section IV, the method is illustrated for a random homogeneous network, the triangular and square lattices. Section V offers a discussion of the results as well as some suggestions for further investigations.

\section{The Model}

Our approach conceives the population as a network, with connections between individuals that do not change in the course of time. Each node of the network represents an individual, and every link symbolizes a relationship between individuals that involves repeated contacts, and therefore the transmission of an infective agent proceeds along connections. Real social networks exhibit rich degree distributions, that is vertices can differ considerably in the number of nearest neighbours. As the aim of this paper is the introduction of a methodology that systematically takes into account higher-order correlations, we adopt the simple susceptible-infected-susceptible model and focus on networks where every node has the same number of nearest neighbours. Despite the homogeneity of these graphs, there exist several classes of such networks differing in topological properties beyond the degree distribution. We shall oppose the regular square lattice to the case where the underlying social structure is fully random, furthermore our approximation scheme is illustrated for a triangular lattice. The generalization to the SIR- or SEIR-models, where the individuals can be in 3 or even 4 possible states, is straightforward.

\begin{figure}[h]
\includegraphics[width=8cm]{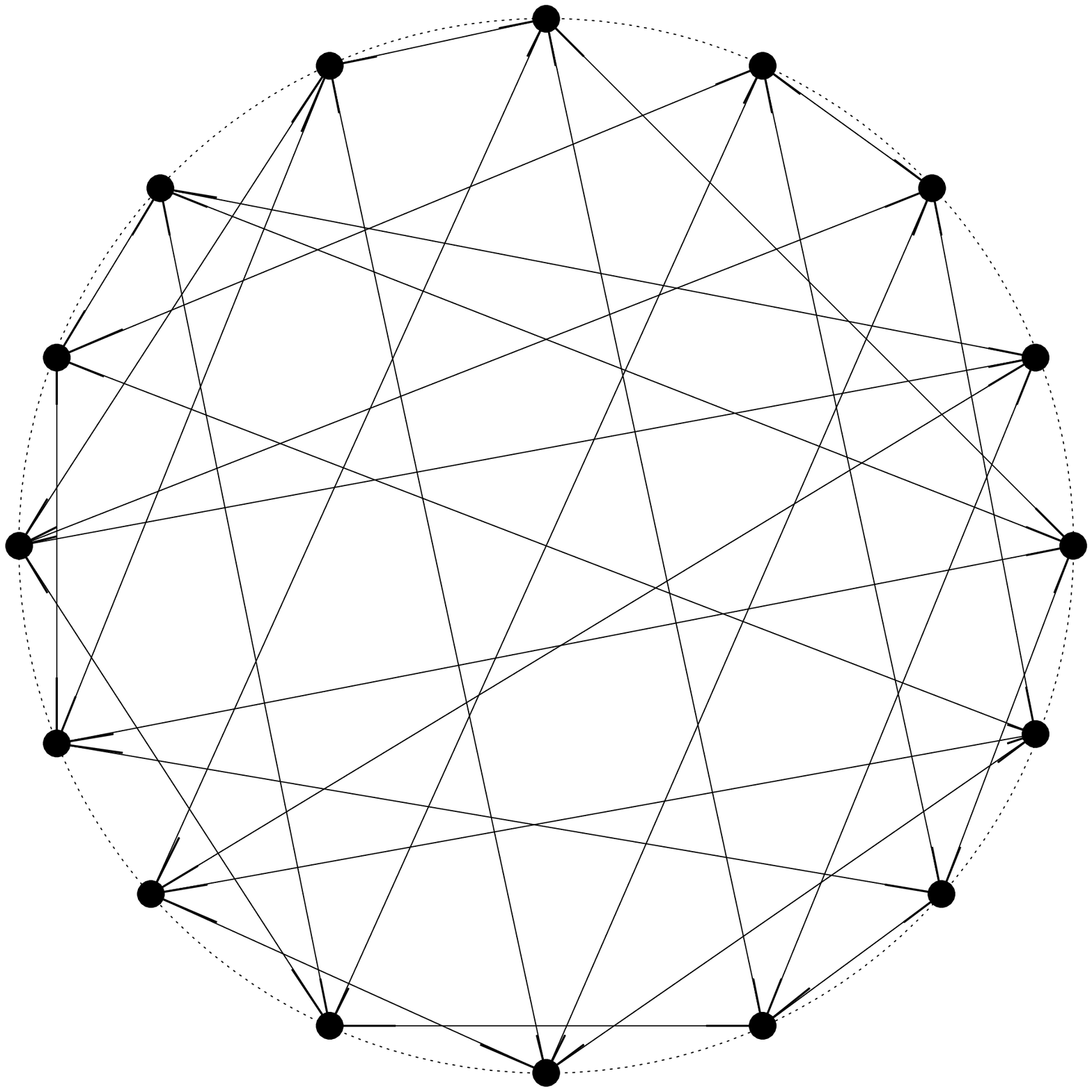}
\caption{Construction of a homogeneous random network of degree $K=4$ and size $N=16$. The nodes are connected randomly and its number of emanating edges are constrained to be $K=4$.}
\label{fig:rand}
\end{figure}

A homogeneous random network of degree $K$ and size $N$ is constructed as follows (Fig.\ \ref{fig:rand}). To each of the $N$ vertices, $K$ ends of edges are attached. The free ends are then connected at random.

The time evolution of the states of the vertices are given by the following rules. Infected nodes recover spontaneously at a rate $\delta$. On the other hand, an infected individual can infect any of its $K$ nearest neighbours at a rate $\nu$. Because what matters is the ratio of the transmission and recovery rates, we can reduce the number of parameters by rescaling the time unit. Thus without loss of generality, the local dynamics is determined by the immunization rate 1 and the effective spreading rate $\lambda=\nu/\delta$. In section III, we will elucidate how this continuous-time model is recovered as a limiting case from a more general discrete-time description.

\section{Revisiting the Mean-Field and Pair Approximations}

In this section, we first review the mean-field and standard pair approximations. These descriptions are obtained by a rate equation which determines the time evolution of some average quantity such as the density of infected individuals or the density of pairs of infected individuals. Up to the level of pair correlations, this is indeed a reasonable approach. But if one wants to keep track of higher-order correlations (for example the density of plaquettes of four infected nodes in the case of the square lattice), a more general starting point may reveal to be advantageous. In subsection B, we derive an exact description of the epidemic dynamics and show how the mean-field and standard pair approximations are recovered in a rather automatic way in part C of this section. The various higher-order approximations are elaborated in the following section.

\subsection{Conventional Approach}

The rate of change of an average quantity $f$ (such as the fraction of sites in a particular state) is described as
\begin{equation} 	\label{equ:aver_rate}
\dot{f}=\sum_{x \in X} \sum_{e_x \in E_x} r(e_x) (f_{e_x}-f),
\end{equation}
where $X$ is the set of all sites, and $E_x$ represents the set of all events that can occur at $x$. A particular event $e_x$ changes the average from $f$ to $f_{e_x}$ and occurs at rate $r(e_x)$ \cite{vanbaalen}.

The SIS-model allows for two possible states, namely susceptible (0) and infected (1). At the mean-field level, the dynamics is described in terms of the density of infected individuals $\rho_1$, and the fraction of susceptible nodes obeys $\rho_0=1-\rho_1$. Eq.\ (\ref{equ:aver_rate}) translates into
\begin{equation}	\label{equ:stand_mf}
\dot{\rho_1}=-\rho_1+\lambda K\rho_0\rho_1.
\end{equation}
The first term accounts for infected nodes becoming healthy whereas the second term describes the new infections, fully ignoring pair correlations.

In the framework of the standard pair approximation \cite{matsuda}, the dynamics is described in terms of the doublet densities $\rho_{xy}$ ($x,y \in \{0,1\}$), this quantity corresponds to the probability that a randomly chosen pair is in configuration $(xy)$. They are related to the global densities $\rho_x$ and local densities (conditional probabilites) $\rho_{x|y}$ by: $\rho_{xy}=\rho_{yx}=\rho_x\rho_{y|x}=\rho_y\rho_{x|y}$. The global and local densities satisfy
$$
\sum_{x=0}^1 \rho_x=1 \quad \text{and} \quad \sum_{x=0}^1 \rho_{x|y}=1 \quad \text{for any} \quad y \in \{0,1\}
$$
Eq.\ (\ref{equ:aver_rate}) tells that the density of infected individuals and the doublet density $\rho_{11}$ evolve in time according to
\begin{equation}	\label{equ:stand_pa}	\begin{split}
\dot{\rho_1} &=-\rho_1+\lambda K\rho_{0|1}\rho_1 \\
\dot{\rho_{11}} &=-2\rho_{11}+2\lambda\rho_{10}+2\lambda(K-1)\rho_{1|01}\rho_{10}.
\end{split}	\end{equation}

\begin{figure}[h]
\includegraphics[width=6cm]{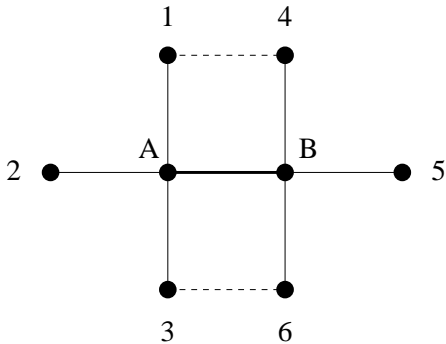}
\caption{An arbitrarily chosen link and its nearest neighbourhood within a homogeneous network characterized by the degree distribution $P(k)=\delta_{k4}$. The dashed lines indicate the connections which are present in the case of a square lattice.}
\label{fig:hom4_pair}
\end{figure}

The first of Eqs.\ (\ref{equ:stand_pa}) can also be regarded as the result of substituting $\rho_0$ by $\rho_{0|1}$ in Eq.\ (\ref{equ:stand_mf}), i.e.\ the susceptible node that is to be infected has to be a nearest neighbour of the vertex which will transmit the infective agent. The second of Eqs.\ (\ref{equ:stand_pa}) includes a recovery term [the first term on the right hand side, destruction of (11)-pairs] and transmission terms [the second and the third terms, creation of (11)-pairs]. The first term describes transitions of pairs in state (11) to either (10) or (01). Both transitions occur at rate 1 (the recovery rate) and thus give rise to the factor 2. The factor 2 in the second and the third terms is needed because we do not assume any asymmetry between sites, which means $\rho_{10}=\rho_{01}$. A (11)-pair can be created from a (10)-pair either if the infective agent proceeds along the connection within that pair (second term) or if the susceptible node is infected by one of the other $K-1$ nearest neighbours of it (third term, see also Fig.\ \ref{fig:hom4_pair}). This path involves the conditional probability $\rho_{1|01}$ [i.e.\ the probability of finding an infected node adjacent to a (01)-pair] which is approximated by $\rho_{1|0}$ as in the ordinary pair approximation, only nearest-neighbour correlations are taken into account. In order to solve the Eqs.\ (\ref{equ:stand_pa}), the system has to be closed. The set $\rho_1, \rho_{1|1}$ is a suitable choice, but $\rho_{11}, \rho_{10}$ works equally well.

\begin{figure}[t]
\includegraphics[width=8cm]{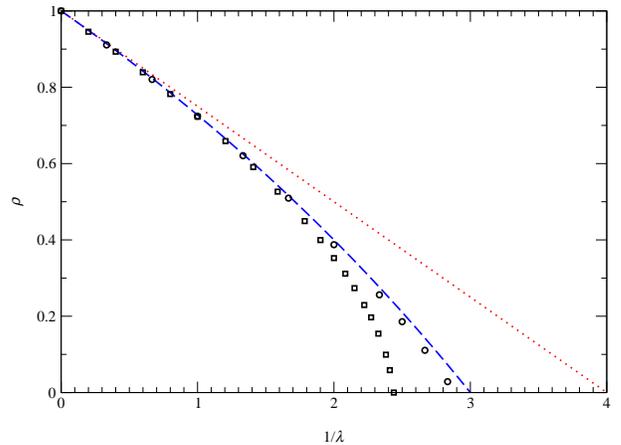}
\caption{Epidemic spreading on homogeneous networks of degree 4. The average number of infected individuals $\rho$ (prevalence) as a function of the inverse infection rate $1/\lambda$ in the steady state is shown. The simulations on the square lattice (squares) and random homogeneous network (circles) exhibit higher epidemic thresholds with respect to the approximations. The mean-field description (dotted line) yields $\lambda_c=1/4$ whereas the pair approximation (dashed line) leads to $\lambda_c=1/3$ for the epidemic threshold. The latter is also in better agreement with the simulation results for $1/\lambda \rightarrow 0$.}
\label{fig:hom4}
\end{figure}

Fig.\ \ref{fig:hom4} contrasts the solutions of Eqs.\ (\ref{equ:stand_mf}) and (\ref{equ:stand_pa}) with the simulations for two different homogeneous networks of degree $K=4$, i.e.\ the square lattice and the one introduced in Fig.\ \ref{fig:rand}. The pair approximation provides a rather good description of the equilibrium dynamics on top of a random homogeneous network, whereas the deviation from the simulation result is remarkable if the population is arranged on a square lattice whose topology is characterized by the presence of many loops of short length. 

We shall now develop a more general formalism that will serve as a starting point in order to investigate the role of correlations beyond the pair level.

\subsection{Exact Description}

In order to arrive at a more general point of departure which will allow us to investigate the role of higher-order spatial correlations, we shall describe the system by assigning a probability $\mathcal P_t({\bf x})$ to every possible configuration ${\bf x}$ at a given time $t$ where each of the $x_i's$ can be either 0 (susceptible) or 1 (infected). This probability is subject to 
$$
\sum_{\bf x} \mathcal P_t({\bf x})=1
$$
at every instant of time. The above introduced SIS-model yields the following transition probabilities for the possible events that can occur at an arbitrary site $l$, involving a discrete time step $\Delta t$
\begin{equation*}	\begin{split}
W_{1 \rightarrow 0}^l &=\Delta t \qquad \qquad W_{0 \rightarrow 0}^l =\prod_{j \text{nn} l} (1-\lambda \Delta t y_j) \\
W_{1 \rightarrow 1}^l &=1-\Delta t \qquad W_{0 \rightarrow 1}^l =1-\prod_{j \text{nn} l} (1-\lambda \Delta t y_j),
\end{split}	\end{equation*}
where the products have to taken over the nearest neighbours of site $l$. By using the binary variables $x_l$ and $y_l$, the above expressions are summarized as 
\begin{equation}	\label{equ:ex_trans}
W_{y_l \rightarrow x_l}^l = x_l+(1-2x_l)\Bigl[\Delta t y_l+(1-y_l)\prod_{j \text{nn} l} (1-\lambda \Delta t y_j)\Bigr].
\end{equation} 
If the total number of nodes is denoted by $N$, the transition probability that the system changes from configuration ${\bf y}$ to ${\bf x}$ can be written as 
\begin{equation}	\label{equ:trans_tot}
\mathcal W_{{\bf y} \rightarrow {\bf x}} = \prod_{l=1}^N W_{y_l \rightarrow x_l}^l,
\end{equation}
and on an exact level, the epidemic dynamics is governed by
\begin{equation}	\label{equ:exact}
\mathcal P_{t+\Delta t}({\bf x}) =\sum_{\bf y} \mathcal W_{{\bf y} \rightarrow {\bf x}} \mathcal P_t({\bf y})
\end{equation}
with $\mathcal W_{{\bf y} \rightarrow {\bf x}}$ given by Eq.\ (\ref{equ:trans_tot}). Eq.\ (\ref{equ:exact}) will serve as starting point for various approximations, be it in discrete or continuous time. In the latter case, only the terms up to order 1 in $\Delta t$ have to be taken into account, but this limit shall be carried out later on. As most of the existing methods are formulated in continuous time, we will elaborate the approximations for this case in order to allow for a comparison.

\subsection{Derivation of the Mean-Field and Pair Approximations}

Within this subsection, it is shown how the approximations (\ref{equ:stand_mf}) and (\ref{equ:stand_pa}) are recovered from the exact description (\ref{equ:exact}).

At the mean-field level, the dynamics is expressed in terms of the density of infected individuals. This quantity corresponds to the probability that an arbitrarily chosen site $i$ is in state $x_i=1$. In order to derive its time evolution, we sum Eq.\ (\ref{equ:exact}) over all possible configurations, $x_i$ held fixed
\begin{equation}	\label{equ:site_1}
\sum_{\{x_j\}_{j \neq i}} \mathcal P_{t+\Delta t}({\bf x})=\sum_{\bf y} \mathcal P_t({\bf y}) \underbrace{\sum_{\{x_j\}_{j \neq i}} \mathcal W_{{\bf y} \rightarrow {\bf x}}}_{W_{y_i \rightarrow x_i}^i}.
\end{equation}
The left hand side of the above equation is $P_{t+\Delta t}(x_i)$, i.e.\ the probability that site $i$ is in state $x_i$ at time $t+\Delta t$. The mean-field approximation consists in considering the sites as independent from each other, i.e.\
\begin{equation}	\label{equ:hom_mix}
\mathcal P_t({\bf y})=\prod_{l=1}^N P_t(y_l),
\end{equation}
which corresponds to the homogeneous mixing hypothesis. Performing the summations, we find for $x_i=1$
\begin{equation}	\label{equ:mf_discr}
P_{t+\Delta t}(1)=1-\Delta tP_t(1)-P_t(0)[1-\lambda \Delta tP_t(1)]^K
\end{equation}
whose continuous-time limit ($\Delta t \rightarrow 0$) is
$$
\dot{P}(1)=-P(1)+\lambda K P(0)P(1),
$$
which is easily identified with Eq.\ (\ref{equ:stand_mf}) since $P(1)=\rho_1$ and $P(0)=\rho_0$.

Let us now see how the pair approximation is obtained by using our formalism. For this purpose, we sum Eq.\ (\ref{equ:exact}) over all possible configurations, $x_A$ and $x_B$ held fixed, where $A$ and $B$ are the two sites of an arbitrarily chosen pair
\begin{equation}	\label{equ:pair_1}
\sum_{\{x_i\}_{i \notin \{A,B\}}} \mathcal P_{t+\Delta t}({\bf x})=\sum_{\bf y} \mathcal P_t({\bf y}) \underbrace{\sum_{\{x_i\}_{i \notin \{A,B\}}} \mathcal W_{{\bf y} \rightarrow {\bf x}}}_{W_{y_A \rightarrow x_A}^A W_{y_B \rightarrow x_B}^B}.
\end{equation}
The left hand side of the above equation corresponds to the probability that the pair $AB$ is in state $(x_Ax_B)$ at time $t+\Delta t$, which shall be denoted by $P_{t+\Delta t}(x_Ax_B)$. By adopting the enumeration introduced in Fig.\ \ref{fig:hom4_pair}, we obtain from Eq.\ (\ref{equ:ex_trans}) for the transition probability $(y_Ay_B) \rightarrow (x_Ax_B)$
\begin{equation}	\label{equ:pair_trans}	\begin{split}
W_{y_A \rightarrow x_A}^A & W_{y_B \rightarrow x_B}^B= \tau_A \tau_B\\
+\Delta t &(1-2x_A)[y_A-\lambda (1-y_A)(y_B+y_1+y_2+y_3)]\tau_B\\
+\Delta t &(1-2x_B)[y_B-\lambda (1-y_B)(y_A+y_4+y_5+y_6)]\tau_A
\end{split}	\end{equation}
where the linearization has been carried out at this point due to technical convenience and
\begin{equation}	\label{equ:tau}
\tau_i=\tau_i(x_i,y_i) \equiv x_i+(1-2x_i)(1-y_i),
\end{equation}
an abbreviation which will also be used below. Furthermore the expression (\ref{equ:pair_trans}) only involves state variables $y_i$ where $i$ is either $A,B$ or one of its nearest neighbours. The sum over the remaining $y_j$ is therefore carried out trivially. Taking into account correlations up to range $2$ only, we write for the probability that the pair $AB$ and its nearest neighbours are in given states
\begin{multline}	\label{equ:hom4pairansatz}
P_t \begin{pmatrix} &y_1&y_4&\\y_2&y_A&y_B&y_5\\&y_3&y_6& \end{pmatrix} = 
P_t(y_Ay_B)\\ 
\times P_t(y_1|y_A)P_t(y_2|y_A)P_t(y_3|y_A)\\
\times P_t(y_4|y_B)P_t(y_5|y_B)P_t(y_6|y_B).
\end{multline}
The conditional probabilities in the above ansatz are expressed as
$$
P(y_i|y_A)=\frac{P(y_iy_A)}{P(y_A)},
$$
where $P(y_A)=\sum_{x=0}^1 P(xy_A)$. Using this ansatz and performing the remaining summations, the continuous-time limit of Eq.\ (\ref{equ:pair_1}) leads to the system (for general $K$)
\begin{equation} 	\label{equ:our_pa}	\begin{split}
\dot{P}(00) &= 2P(10)\left[1-\lambda(K-1)\frac{P(00)}{P(0)}\right] \\
\dot{P}(10) &= P(11)-P(10)+\lambda P(10)\left[2(K-1)\frac{P(00)}{P(0)}-K\right]\\
\dot{P}(11) &= -2P(11)-2\lambda P(10)\left[(K-1)\frac{P(00)}{P(0)}-K\right].
\end{split}	\end{equation}
By identifying the pair probabilites $P(xy)$ with the doublet densities $\rho_{xy}$ and since $\rho_{00}/\rho_0=1-\rho_{10}/\rho_0$, the system of Eqs.\ (\ref{equ:our_pa}) corresponds to Eqs.\ (\ref{equ:stand_pa}).

In summary, in the standard derivation of the mean-field and pair approximations based on Eq.\ (\ref{equ:aver_rate}), the rate of change of an average density is directly expressed by all the different events that can alter its value in a rather heuristic way [Eqs.\ (\ref{equ:stand_mf}) and (\ref{equ:stand_pa})]. On the other hand, the derivation of the approximations becomes an automatic procedure involving
\begin{itemize}
\item{an initial summation of the system probability $\mathcal P_{t+\Delta t}({\bf x})$ over all possible states except a few in order to obtain $P_{t+\Delta t}(x)$ or $P_{t+\Delta t}(x_Ax_B)$ [Eqs.\ (\ref{equ:site_1}) and (\ref{equ:pair_1})],}
\item{an ansatz corresponding to the approximation [Eqs.\ (\ref{equ:hom_mix}) and (\ref{equ:hom4pairansatz})],}
\item{and the continuous-time limit.}
\end{itemize}
However, the last step is not really imperative. Our methodology works equally well in discrete time. If $\Delta t$ is set to 1, $\lambda \Delta t=\lambda$ then corresponds to a probability rather than to a rate and higher-order terms in $\lambda$ appear in the equations. As an example, the discrete-time evolution at the mean-field level is governed by Eq.\ (\ref{equ:mf_discr}). Obviously, the results quantatively differ from the continuous-time limit. The full advantage of this formalization will become clear in the next section.

It is also important to note that topological properties beyond the degree distribution do not enter at the level of the standard pair approximation. In the case of the square lattice, the nodes 1 and 4 as well as 3 and 6 (Fig.\ \ref{fig:hom4_pair}) are also connected whereas these links are missing in its random counterpart. Various improvements upon the ordinary pair approximation have been proposed. Instead of deriving the higher-order correlations from the dynamics of the system, these pair models consist in making a number of biologically motivated assumptions involving parameters that characterize the topology of the underlying network. We shall compare our approach with these improved pair models in the next section.

\section{Further Systematic Improvement}

The difference between the simulation result and the pair approximation in Fig.\ \ref{fig:hom4} is rooted in the neglection of correlations of range greater than $2$. Especially in the vicinity of the phase transition, where a finite fraction of the nodes starts being infected, the links should not be considered independently, and higher-order dynamical correlations have to be taken into account. In other words, the state $x_i$ of node $i$ at time $t+\Delta t$ is determined by all the states of its nearest neighbours of node, i.e.\ it is not the case that the states of the various nearest neighbours at time $t$ contribute independently from each other to the state $x_i$ at time $t+\Delta t$.

We therefore want to incorporate the longer correlation range by extending the fundamental cluster (site, pair) to a star or square, respecting the underlying network's topology. Therefore different spatial patterns are embedded very naturally by our method. The dynamical equations, to which the higher-order correlations are subject to, are derived in a very straightforward way by our formalism. The binary nature of these equations allows for a very efficient  solution by the computer. On the other hand, the equations can be simplified further by taking into account the underlying symmetries. This latter procedure will be illustrated for the triangular and square lattices. Performing this extension, we find an improved description of the steady state as well as the dynamics.

Alternatively, it is possible to derive the dynamics of triple correlations by using Eq.\ (\ref{equ:aver_rate}). Although this approach has the advantage that no specific cluster must be chosen, it is a rather difficult undertaking \cite{morris}.

\subsection{Homogeneous Random Network}

Random networks are characterized by a vanishing clustering (local interconnectedness), but the average distance between any pair of nodes only increases logarithmically with the system size: this is known as the small world phenomenon \cite{watts}. It is easy to imagine that the epidemic spreads the more rapidly the smaller the underlying ``world'' is.

\begin{figure}[t]
\includegraphics[width=6cm]{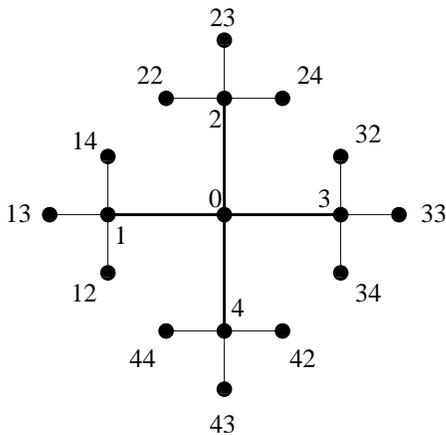}
\caption{An arbitrary node (denoted by 0) with its corresponding starlike fundamental cluster within a homogeneous random network of degree $K=4$.}
\label{fig:enum_star}
\end{figure}

As the local topology is fully treelike, we shall use a star as our fundamental element. In contrast to regular lattices, this extension is a unique choice. Fig.\ \ref{fig:enum_star} shows an arbitrarily chosen node in a homogeneous random network and two hierarchies of its nearest neighbours, also introducing the notation which is adopted below.

The probability that, at time $t$, node $0$ is in state $x_0$ and its nearest neighbours $1,2,3,4$ are in the states $\{x_1,x_2,x_3,x_4\}$ is denoted by 
$$
P_t \begin{pmatrix} &x_2& \\ x_1&x_0&x_3 \\ &x_4& \end{pmatrix}
$$ 
and obtained by summing $\mathcal P_t({\bf x})$ over all possible configurations, $\{x_0,x_1,x_2,x_3,x_4\}$ being fixed.

The probability that the nearest and second-nearest neighbours of node $0$ are in given states, is given by the ansatz (the sum over the remaining y-states is again performed trivially)
\begin{equation}	\label{equ:star_ansatz}	
P_t(\{y_j\}_{j \in \mathcal N}) = P_t \begin{pmatrix} &y_2& \\ y_1&y_0&y_3 \\ &y_4& \end{pmatrix}  \prod_{l=1}^4 P_t(y_{l2}y_{l3}y_{l4}|y_ly_0),
\end{equation}
where $\mathcal N$ represents the set of nodes depicted in Fig.\ \ref{fig:enum_star} and the conditional probabilities are
$$
P_t(y_{l2}y_{l3}y_{l4}|y_ly_0)=\frac{P_t \begin{pmatrix} &y_{l4}& \\ y_{l3}&y_l&y_0 \\ &y_{l2}& \end{pmatrix} }{P_t(y_ly_0)}.
$$
The pair probability appearing in the above expression is extracted from the corresponding star probabilities by 
$$
P_t(y_ly_0)=\sum_{y_{l2}=0}^1 \sum_{y_{l3}=0}^1 \sum_{y_{l4}=0}^1 P_t \begin{pmatrix} &y_{l4}& \\ y_{l3}&y_l&y_0 \\ &y_{l2}& \end{pmatrix} .
$$ 

\begin{figure}[h]
\includegraphics[width=9cm]{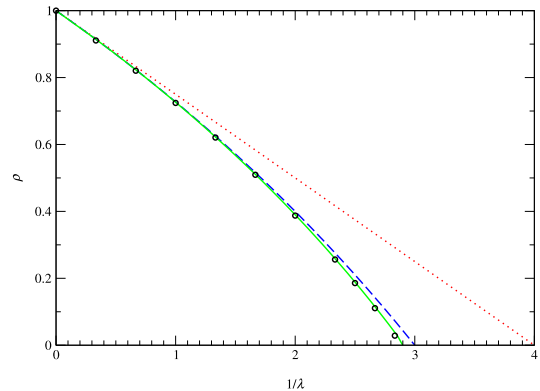}
\caption{Equilibrium prevalence of the epidemic process on a homogeneous random network with $P(k)=\delta_{k4}$. The star approximation (solid line) is in excellent agreement with the simulation result (circles), yielding also an accurate description of the critical region. The mean-field (dotted line) and pair approximation (dashed line) have been plotted again for comparison.} 
\label{fig:hom_res}
\end{figure}

\begin{figure}[t]
\includegraphics[width=8cm]{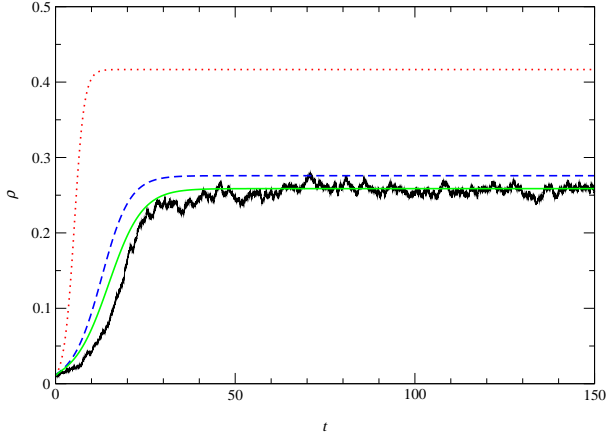}
\caption{Invasion of an infective agent (infection rate $\lambda=3/7$) in a population whose connectivity patterns are given by a homogeneous random network. At the mean-field level (dotted line), the initial prevalence of 0.01 increases to its equilibrium value during 10 time units only. The pair approximation (dashed line) provides a further improvement, and the star approximation (solid line) is in remarkable agreement with the stochastic simulation.} 
\label{fig:hom_dyn}
\end{figure}

With these ingredients, the continuous-time limit of Eq.\ (\ref{equ:exact}) reads
\begin{equation}	\label{equ:star_approx}	\begin{split}	
\dot{P}\begin{pmatrix} &x_2& \\ x_1&x_0&x_3 \\ &x_4& \end{pmatrix} &=\sum_{\{y_j\}_{j \in \mathcal N}} [  P(\{y_j\}_{j \in \mathcal N}) \\
\prod_{i=0}^4 &\left\{(1-2x_i)[y_i-\lambda(1-y_i)\sum_{j\text{nn}i}y_j]\prod_{k \neq i} \tau_k\right\}]
\end{split}	\end{equation}
with $P(\{y_j\}_{j \in \mathcal N})$ given by Eq.\ (\ref{equ:star_ansatz}). The binary character of this system of $2^5=32$ equations permits a very efficient numerical implementation. On the other hand, if one takes into account the symmetries of the problem, the degrees of freedom can be reduced to 10, but this procedure will be shown for the regular lattices. Fig.\ \ref{fig:hom_res} shows the striking agreement of the star approximation with the simulation result, all along from a high effective spreading rate to its threshold value, for the equilibrium situation. Fig.\ \ref{fig:hom_dyn} opposes the various approximations to the stochastic simulation for the case of the invasion of an infective agent, the initial prevalence being set to 0.01. Whereas the steady state is reached rather quickly in the mean-field description, the slope of the star approximation is in remarkable agreement with the simulation. As correlations of a greater range are taken into account, it can also be observed that the system equilibrates more smoothly, that is $\ddot{\rho}(t \simeq 30)$ for the star approximation is considerably smaller than the rate of change of $\dot{\rho}$ at time $t \simeq 10$ at the mean-field level.

\subsection{Square Lattice}

\begin{figure}[h]
\includegraphics[width=5cm]{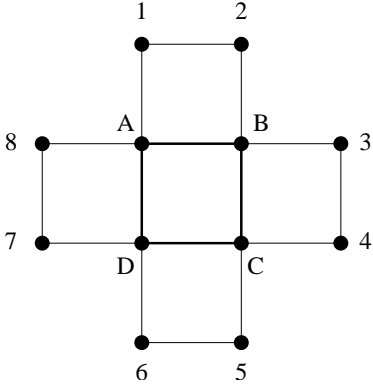}
\caption{An arbitrarily chosen square within a 2-dimensional lattice and the denotation of the nearest neighbours of its corners. The former serves as the fundamental element within the square approximation.}
\label{fig:sq_sq}
\end{figure}

In contrast to random graphs, the epidemic dynamics on top of this regular network is essentially dominated by the presence of loops. In order to arrive at a level of description beyond the pair approximation, we shall use the square as our fundamental cluster. This seems to be a natural choice, although it is not unique as discussed below. In analogy to the previous subsection, the probability that the corners of the square $ABCD$ are in the states $\{x_A,x_B,x_C,x_D\}$ at time $t$ is deduced from the system probability by
$$
P_t \begin{pmatrix} x_A&x_B\\x_D&x_C \end{pmatrix} =\sum_{\{x_i\}_{i \notin \{A,B,C,D\}}} \mathcal P_t({\bf x}).
$$
If the nearest neighbours of the vertices $A,B,C$ and $D$ are enumerated according to Fig.\ \ref{fig:sq_sq}, we write for the probability that the nodes comprised within these 5 squares (i.e.\ the nearest neighbours of the central plquette) are in given states
\begin{multline}	\label{equ:sq_prob}
P_t(\{y_i\}_{i \in \{A,B,C,D,1,2,...,8\}})= P_t\begin{pmatrix} y_A&y_B\\y_D&y_C\end{pmatrix} \\
\times P_t(y_1y_2|y_Ay_B)P_t(y_3y_4|y_By_D)\\
\times P_t(y_5y_6|y_Cy_D)P_t(y_7y_8|y_Ay_C)
\end{multline}
with
$$
P_t(y_1y_2|y_Ay_B)=\frac{P_t\begin{pmatrix}y_1&y_2\\y_A&y_B\end{pmatrix}}{P_t(y_Ay_B)}
$$
involving the pair probability 
$$
P_t(y_Ay_B)= \sum_{x_1=0}^1 \sum_{x_2=0}^1 P_t\begin{pmatrix}x_1&x_2\\y_A&y_B\end{pmatrix}
$$
and analogously for the other factors appearing in (\ref{equ:sq_prob}). At this point, we could again write down an equation of the type (\ref{equ:star_approx}), but the $2^4=16$ plaquette probabilities are subject to various symmetries, which we will denote by

\begin{figure}[t]
\includegraphics[width=10cm]{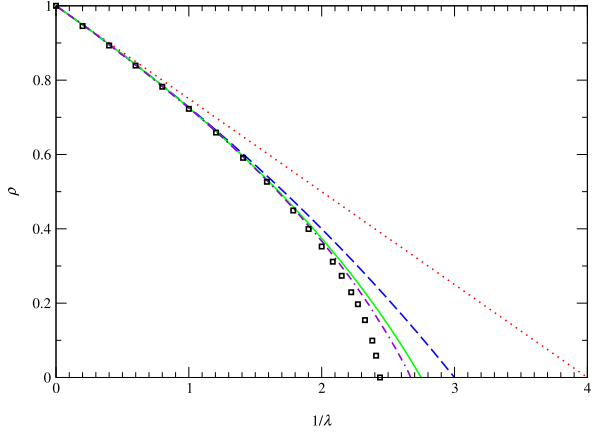}
\caption{Equilibrium prevalence in a square lattice structured population. The mean-field (dotted line) and pair approximations (dashed line) are levels of description at which topological properties beyond the degree distribution do not enter. The approximations involving the square (solid line) and a rectangle composed of two squares (dashed-dashed-dotted line) as fundamental units are shown to yield a systematic improvement of the steady-state behaviour.}
\label{fig:sq_imp}
\end{figure} 

\begin{equation*}	\begin{split}
P_t{00 \choose 00} &\equiv q_{0,t}\\
P_t{10 \choose 00} &=P_t{01 \choose 00}=P_t{00 \choose 10}=P_t{00 \choose 01} \equiv q_{1,t}\\
P_t{11 \choose 00} &=P_t{01 \choose 01}=P_t{00 \choose 11}=P_t{10 \choose 10} \equiv q_{2,t}^A\\
P_t{10 \choose 01} &=P_t{01 \choose 10} \equiv q_{2,t}^C\\
P_t{11 \choose 10} &=P_t{11 \choose 01}=P_t{10 \choose 11}=P_t{01 \choose 11} \equiv q_{3,t}\\
P_t{11 \choose 11} & \equiv q_{4,t}
\end{split}	\end{equation*}
The exact description (\ref{equ:exact}) leads to the following continuous-time dynamics for these quantities
\begin{equation} 	\begin{split}	\label{equ:sq_appr}	
\dot{q}_0 &=4q_1-8\lambda T_1q_0\\
\dot{q}_1 &=-q_1+2q_2^A+q_2^C+\lambda[-2q_1(1+2T_1+T_2)+2T_1q_0]\\
\dot{q}_2^A &=-2q_2^A+2q_3+\lambda[-4q_2^A+2T_1(q_0+3q_1)+2T_2(q_1-q_2^A)]\\
\dot{q}_2^C &=-2q_2^C+2q_3+\lambda(-4q_2^C+4T_1q_1-4T_2q_2^C)\\
\dot{q}_3 &=-3q_3+q_4+\lambda[2q_1+4q_2^A-4q_3 -2T_1(q_0+2q_1)\\
&+2T_2(q_1+2q_2^A+2q_2^C)]\\
\dot{q}_4 &=-4q_4+\lambda[8q_2^C+16q_3-8T_2(q_1+q_2^A+q_2^C)]
\end{split}		\end{equation}
where
$$
T_1=\frac{t_1^A}{p_0^A} \qquad \text{and} \quad T_2=\frac{t_2^C}{p_1^A}
$$
involving the following triplet- and pair probabilites given by the square probabilities through
\begin{equation*}		\begin{split}	
t_1^A &= P\begin{pmatrix} 0&0\\ 1&\end{pmatrix}=q_1+q_2^A, \qquad t_2^C=P\begin{pmatrix}0&1\\1&\end{pmatrix}=q_2^C+q_3,\\
p_0^A &= P(00)=q_0+2q_1+q_2^A \qquad \text{and}\\
p_1^A &=P(10)=q_1+q_2^A+q_2^C+q_3.
\end{split}	\end{equation*}
Since
$$
q_0+4q_1+4q_2^A+2q_2^C+4q_3+q_4=1,
$$
the square approximation in the form (\ref{equ:sq_appr}) represents a dynamical system of 5 degrees of freedom. It also has to be noted that the computational load of the square approximation in this form is reduced dramatically with respect to its raw form [analogous version of Eq.\ (\ref{equ:star_approx})].

\begin{figure}[h]
\includegraphics[width=8cm]{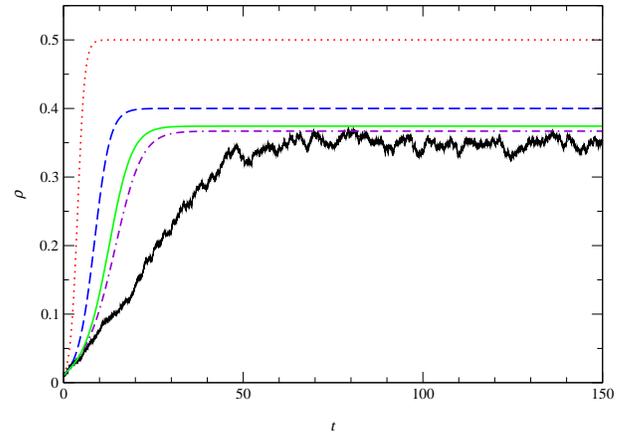}
\caption{Epidemic dynamics in a square lattice structured population (transmission rate $\lambda=1/2$). By taking into account correlations of a greater range, the slope during the transient time decreases as a comparison of the mean-field (dotted line), pair (dashed line), square (solid line) and bisquare approximations (dashed-dashed-dotted line) shows. The difference between the simulation result and the bisquare approximation remains significant during the invasion period due to the considerable effect of random events at overall low prevalence.} 
\label{fig:sq_dyn}
\end{figure}

Fig.\ \ref{fig:sq_imp} shows the systematic improvement brought about by the square- and the bisquare approximations in dynamical equilibrium. The latter is a description whose fundamental cluster is composed of two squares. Its prediction of the epidemic threshold ($\lambda_c \simeq 0.38$) is still lower than the simulation result ($\lambda_c \simeq 0.41$): this highlights the crucial role of the higher-order spatial correlations in lattice structured populations. Fig.\ \ref{fig:sq_dyn} represents the improvements upon the dynamics. Note that from a certain characteristic time the simulation lags behind all the approximations as a direct consequence of the stochasticity which is particularly important at low prevalences. However this characteristic time is shifted to the right as higher-order correlations of a greater range are taken into account.

An improvement upon the standard pair approximation can also be obtained as follows \cite{vanbaalen}. Instead of deriving the square probabilities from the dynamics of the system, one can write it as 
$$
P\begin{pmatrix}x_i&x_a\\x_j&x_b\end{pmatrix}=P(x_i)P(x_a)P(x_b)P(x_j)C_{ia}C_{ab}C_{bj}C_{ji}T_{\Box iabj}
$$
involving the relative pair and square correlation factors $C_{xy}$ and $T_{\Box iabj}$. For a straight triple, it is supposed
$$
P(x_ix_ax_b)=P(x_i)P(x_a)P(x_b)C_{ia}C_{ab}T_{\angle iab}.
$$
By setting the relative correlation factors $T_{\Box iabj}$ and $T_{\angle iab}$ to 1 and using the fact that on the square lattice $1/3$ of the triples are straight and $2/3$ form part of a square, one obtains an improvement for $P(x_i|x_ax_b)$. In other words, the conditional probability $P(x_i|x_ax_b)$ is not simply set to $P(x_i|x_a)$ as it is done in the ordinary pair approximation, but rather the loop structure is incorporated while still using pairs as building blocks.

\subsection{Triangular Lattice}

\begin{figure}[t]
\includegraphics[width=4cm]{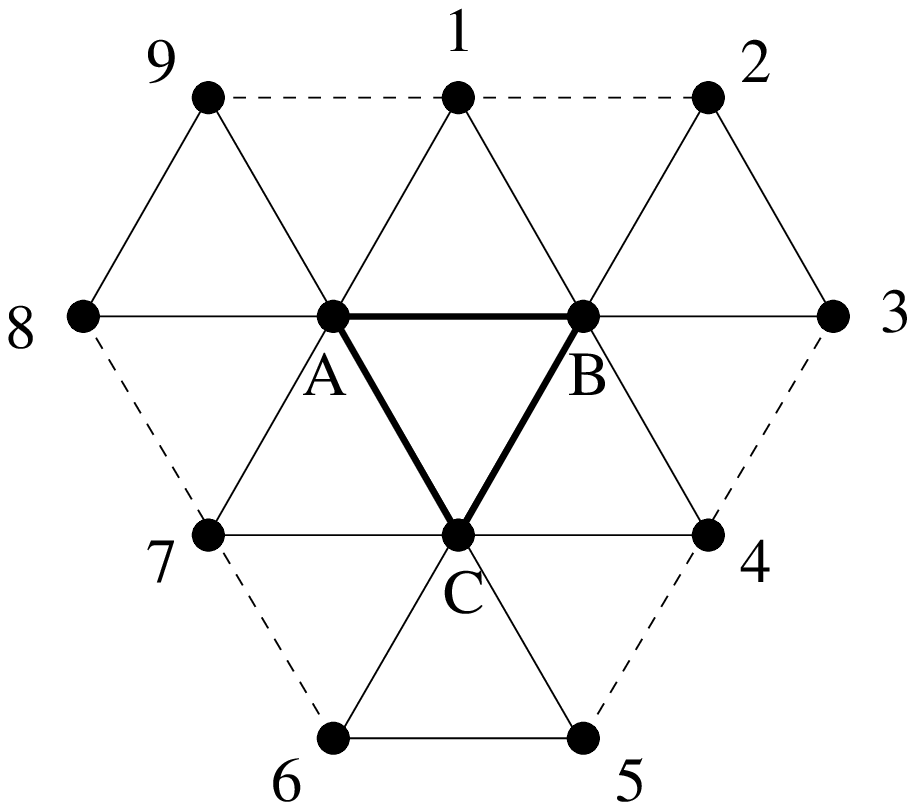}
\caption{An arbitrarily chosen triangle and its nearest neighbourhood. The dashed lines indicate that the corresponding links are ignored.}
\label{fig:tri_denot}
\end{figure}

In its ordinary formulation, the fact that two sites can have neighbours in common, is simply ignored by the pair approximation. By means of the triangular lattice, we show how the method introduced in this paper has to be applied, i.e. what the next level of description beyond the pair approximation is.

The clue is to use the triangle as the basic element. In analogy to the previous cases, the probability that the vertices of a triangle $ABC$ are in the states $\{x_A,x_B,x_C\}$ at time $t$ is obtained through
$$
P_t{x_Ax_B\choose x_C}=\sum_{\{x_i\}_{i \notin \{A,B,C\}}} \mathcal P_t({\bf x}).
$$
Fig. \ref{fig:tri_denot} shows the neighbourhood of an arbitrarily chosen triangle within this lattice. For the probability that the vertices depicted in Fig.\ \ref{fig:tri_denot} are in given states, we suppose
\begin{equation*}	\begin{split}
P_t(\{y_i\}_{i \in \{A,B,C,1,2,...,9\}})= P_t{y_Ay_B \choose y_C} \\
\times P_t(y_1|y_Ay_B)P_t(y_4|y_By_C)P_t(y_7|y_Cy_A) \\
\times P_t(y_8y_9|y_A)P_t(y_2y_3|y_B)P_t(y_5y_6|y_C).
\end{split}	\end{equation*}
The conditional probabilities appearing in the above expression can be written as fractions involving site- and pair probabilities. The latter are deduced from the triangle probabilities in analogy to previous explanations. As the triangle correlations are subject to the symmetries
\begin{equation*}	\begin{split}
P_t{00\choose 0} &\equiv t_{0,t}\\
P_t{10\choose 0} &= P_t{01\choose 0} = P_t{00\choose 1} \equiv t_{1,t}\\
P_t{11\choose 0} &= P_t{10\choose 1} = P_t{01\choose 1} \equiv t_{2,t}\\
P_t{11\choose 1} &\equiv t_{3,t},
\end{split}	\end{equation*}
a further simplification can be performed, and finally the continuous-time triangle dynamics is governed by the equations
\begin{equation}	\begin{split}	\label{equ:tri_appr}
\dot{t}_0 &= 3[t_1-2\lambda(A_1+A_2)]\\
\dot{t}_1 &= -t_1+2t_2+2\lambda(-2t_1+3A_1+2A_2-2A_3-A_4)\\
\dot{t}_2 &= -2t_2+t_3+2\lambda(t_1-3t_2-3A_1-A_2+4A_3+2A_4)\\
\dot{t}_3 &= 3[-t_3+2\lambda(t_1+3t_2+A_1-2A_3-A_4)].
\end{split}	\end{equation}
where
\begin{equation*}	\begin{split}
A_1 &= \frac{p_1t_0}{s_0}, \qquad A_2=\frac{t_0t_1}{p_0}\\
A_3 &= \frac{p_0p_1}{s_0} \quad \text{and} \quad A_4=\frac{t_1t_2}{p_1}
\end{split}	\end{equation*}
depending on the pair probabilities $p_1=P(10)=t_1+t_2$, $p_0=P(00)=t_0+t_1$ and the site probability $s_0=P(0)=t_0+2t_1+t_2$.

Because of the constraint
$$
t_0+3t_1+3t_2+t_3=1,
$$
we have three degrees of freedom in the triangle approximation (\ref{equ:tri_appr}).

\begin{figure}[t]
\includegraphics[width=10cm]{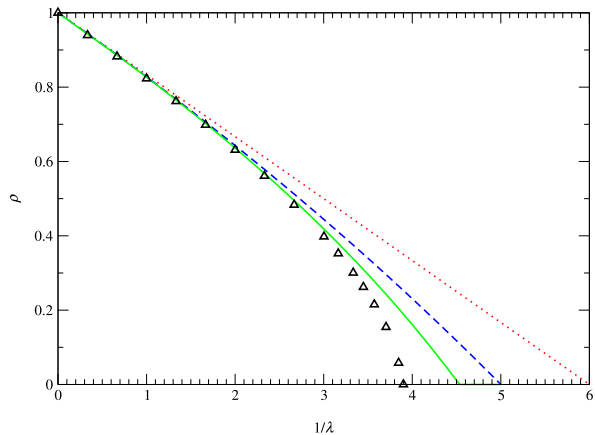}
\caption{Steady-state prevalence in a triangular lattice structured population. The mean-field description (dotted line) yields an epidemic threshold $\lambda_c=1/6$. With respect to the pair approximation (dashed line), the description based on the triangle (solid line) provides a better approximation of the simulation result (triangles).}
\label{fig:tri_res}
\end{figure}

\begin{figure}[h]
\includegraphics[width=8cm]{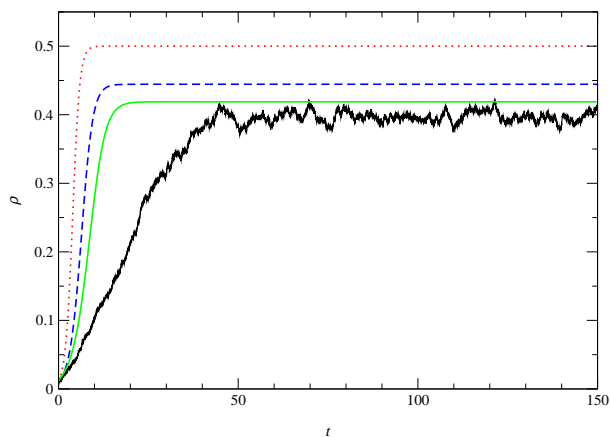}
\caption{Invasion dynamics in a population represented by a triangular lattice ($\lambda=1/3$). The upper two curves show the mean-field (dotted line) and pair dynamics (dashed line). The improvement brought about by the triangle approximation (solid line) still lags behind the simulation result due to the same reason as in the case of the square lattice.} 
\label{fig:tri_dyn}
\end{figure}

As far as the equilibrium prediction is concerned, the triangle approximation provides a very good description for $1/\lambda < 3$ (Fig.\ \ref{fig:tri_res}). The difference between its threshold prediction ($1/\lambda_c \simeq 4.5$) and the simulation result ($1/\lambda_c \simeq 3.9$) is of the same order of magnitude as the plaquette approximation in the case of the square lattice. Concerning the dynamics (Fig.\ \ref{fig:tri_dyn}), we also observe a lag between the simulation and the approximations, and the slope during the transient time is slightly improved as one goes from the pair to the triangle approximation.

The strategy outlined at the end of the last subsection can also be applied to the triangular lattice \cite{vanbaalen}. In addition to the open triplet probability, the triangle probability is written as
$$
P{x_i\choose x_a x_b}=P(x_i)P(x_a)P(x_b)C_{ia}C_{ab}T_{\triangle iab}.
$$
One then obtains an analogous correction for $P(x_i|x_ax_b)$ involving a parameter $\theta$ denoting the fraction of triplets in closed form which is $2/5$ in the triangular lattice. Interestingly, the simplest elaboration of this approach ($\tau_{\triangle iab}=\tau_{\angle iab} \equiv \tau_{iab}$) reproduces the invasive period reasonably accurate if $\theta$ is chosen larger than its correct value ($\theta \simeq 0.6$). Keeling et al.\ and Rand also developed improved pair models based on this approach \cite{keeling97,rand}.

\section{Conclusion}

We have studied a dynamical model of epidemic spreading where every individual is in contact with an equal number $K$ of nearest neighbours. Infected nodes recover spontaneously at a rate 1, on the other hand, they infect neighbouring susceptible sites at a rate $\lambda$. We have chosen this simple SIS-type model since the focus of this article is the introduction of a novel methodology that allows a rather straightforward derivation of the dynamics of higher-order correlations.

The method we propose here consists in choosing a fundamental cluster composed of a certain number of nodes $n$ as well as links connecting them. A definite probability is assigned to each possible configuration of the basic element. The size of the fundamental cluster represents the range up to which spatial correlations are exactly taken into account. At a level beyond the pair approximation, the choice of the basic element is guided by the underlying network's topology. In the case of the square lattice, clusters composed of at least one plaquette serve as the fundamental element; for random networks the local treelike structure is incorporated by using the star as the basic unit. Spatial patterns beyond the degree distribution are therefore embedded in a very natural way by our method. Describing the epidemic dynamics of the entire population as a discrete time Markovian process, the appearing probabilities (probability that a cluster and its nearest neighbourhood is in a given configuration) are expressed in terms of the fundamental cluster probabilities. The continuous-time dynamics emerges as a limiting case ($\Delta t \rightarrow 0$).

With respect to the ordinary (rather heuristic) derivation of the mean-field and pair approximation, these descriptions are derived with the help of our formalism by using the site or the pair respectively as fundamental clusters in a very automatic way. Independently of the specific choice of the cluster, the binary character of the resulting equations allows for a very efficient solution by the computer. Likewise, a further simplification can be reached if the symmetries which the fundamental cluster probabilites are subject to, are taken into account. As soon as correlations of range greater than $2$ are not ignored, our method yields improved estimates for the location of the phase transition. In the case of the random network, the star approximation already leads to an excellent description of the steady state and the transient dynamics. In the regular counterpart, many squares have to be included within the corresponding fundamental unit in order to attain the same level of accuracy. This is due to the presence of stronger correlations caused by the high local networking. The method was also illustrated for a triangular lattice and contrasted to approaches that make a certain number of assumptions about the higher-order correlations which lead to improved pair models.

However the novelty of the present work lies in the formalism which essentially consists in a more general starting point and its associated systematic improvability rather than the specific results for the selected epidemiological model and geometrical examples.

\section{Acknowledgment}
We wish to thank EC-Fet Open project COSIN IST-2001-33555, and the OFES-Bern (CH) for financial support.

\end{document}